\documentclass[a4paper]{jpconf}
\usepackage{graphicx}

\newcommand{\beq}{\begin{equation}}
\newcommand{\eeq}{\end{equation}}
\newcommand{\pp}{\partial}
\newcommand{\lap}{\nabla^2}
\newcommand{\lapi}{\nabla^{-2}}
\newcommand{\benn}{\begin{eqnarray}}
\newcommand{\eenn}{\end{eqnarray}}

\begin{document}
\title{Gyrofluid simulations of collisionless reconnection in the presence of diamagnetic effects}

\author{E Tassi$^1$, F L Waelbroeck$^2$ and D Grasso$^{3,4}$}

\address{$^1$Centre de Physique Th\'eorique, CNRS -- Aix-Marseille Universit\'es, Campus de Luminy, case 907, F-13288 Marseille cedex 09, France}
\address{$^2$Institute for Fusion Studies, The University of Texas at Austin, Austin, TX 78712-1060, USA}
\address{$^3$ Istituto dei Sistemi Complessi - CNR, Roma, Italy}
\address{$^4$ Dipartimento di Energetica, Politecnico di Torino, Italy}

\ead{tassi@cpt.univ-mrs.fr}

\begin{abstract}

The effects of the ion Larmor radius on magnetic reconnection are investigated by means of numerical simulations, with a Hamiltonian gyrofluid model. In the linear regime, it is found that ion diamagnetic effects decrease the growth rate of the dominant mode. Increasing ion temperature tends to make the magnetic islands propagate in the ion diamagnetic drift direction.  In the nonlinear regime, diamagnetic effects reduce the final width of the island. Unlike the electron density, the guiding center density does not tend to distribute along separatrices and at high ion temperature, the electrostatic potential exhibits the superposition of a small scale structure, related to the electron density, and a large scale structure, related to the ion guiding-center density. 

\end{abstract}

\section{Introduction}

Magnetic reconnection is responsible for many phenomena occurring in laboratory and astrophysical plasmas. \cite{Pri00,Bis00} In this paper, we restrict consideration to reconnection in low-beta configurations (i.e. $m_e/M_i < \beta \ll 1$, with $m_e$ and $M_i$ indicating electron and ion mass, respectively ) characterized by the presence of a magnetic ``guide'' field perpendicular to the reconnection plane. The guide field ensures that the ions undergo rapid Larmor gyration as they cross the reconnection region.  Guide field reconnection is thought to be responsible, in particular, for the sawtooth crash in tokamaks. Kinetic simulations of guide-field reconnection are generally more costly and difficult than in the alternative case of ``component reconnection,'' due to the separation of scales imposed by the guide field. \cite{WessonPPCF,Udi05,LazWaeLuce} In the linear \cite{CoppiGal} as well as in the nonlinear regime, \cite{Jribbon,Zak93,RogrsZakh96} magnetic reconnection is characterized by processes occurring in very narrow resonant layers. For high-temperature plasmas, such as those existing in the present generation of tokamaks, the resonant layer is narrower than the ion Larmor radius, so that the ions may traverse the entire layer during a cyclotron period.\cite{drake-kinIK} Their response is then a nonlocal functional of the fields they sample along their orbit. \cite{Por91,Berk91}

Conventional fluid models treat finite Larmor radius (FLR) effects by expanding the ion response to first order in $(k_\bot\rho_i)^2$. \cite{Bisk81,HKM85,Bis97,FitzWatsWael} In the linear regime, the predictions of fluid models that include FLR effects \cite{Zak2flu92} are qualitatively similar to the results of kinetic calculations \cite{Por91,Berk91} that account for the nonlocal response of the ions. In particular, both fluid and kinetic models predict the stabilization of ideal instabilities when the growth rate $\gamma_0$ computed in the absence of diamagnetic drifts is less than half of $\omega_{*i}$, the ion diamagnetic drift angular frequency. \cite{Por91,Zak2flu92,Huy01} It is unclear {\em a priori}, however, whether fluid FLR calculations such as the investigations of the nonlinear saturation of reconnecting modes in Refs.~\cite{Bisk81,Bis97,RogZakhomstar}, offer an adequate description of ion dynamics in the nonlinear regime. In particular, it is unclear whether they are adequate for the task of extrapolating the experimental observations to different values of $\rho_i/L$, where $L$ describes the plasma size. Some indications are provided by nonlinear simulations using gyrokinetic codes, \cite{Matsu99} but numerical limitations make scaling studies with such codes impractical. 

The need for a simpler nonlinear model for the nonlocal ion response has motivated the derivation and investigation of gyrofluid models that account for these physical effects in a fluid context. Grasso {\em et al.} \cite{Gra00} introduced the first electromagnetic gyrofluid model and applied it to the study of magnetic reconnection. Their gyrofluid model, although unable to describe kinetic phenomena such as wave particle resonance (Landau damping), \cite{Berk91} nevertheless illuminated the qualitative properties of the reconnection dynamics. More elaborate models that do account for Landau damping were subsequently developed by Snyder and Hammett, but the applications of their models have been limited to the study of electromagnetic effects on turbulence driven by pressure gradients. \cite{SnydHamm01} 

In the present contribution, we investigate a recently derived 3-field gyrofluid model \cite{Wae09} by means of numerical simulations. This gyrofluid model is part of a family of Hamiltonian fluid models  \cite{Caf98,Gra00,Tas08,Tas10} that have been investigated in past years, but is distinguished from these other Hamiltonian models by its nonlocal treatment of the ion density evolution. In the linear regime, the nonlocal description of the ion density evolution improves on FLR models by describing the evanescence of drift waves excited at frequencies in the ion drift direction but below the diamagnetic frequency. \cite{Wae09} The propagation and evanescence properties of drift waves can be important for determining the role of the polarization current in island evolution. \cite{FitzWaelWave05} 

In recent work, \cite{Gra10} we have used the 3-field model of Ref.~\cite{Wae09} to revisit earlier investigations \cite{Gra00} of gyrofluid magnetic reconnection in a {\em homogeneous} plasma, where ion diamagnetic effects vanish. The present article extends these studies to the case of {\em inhomogeneous} plasma, where ion diamagnetic drifts affect the reconnection. Such ion drift effects can be important for sawtooth dynamics in tokamaks \cite{RogrsZakh96,Bisk81,Bis97}, where it accounts for the saturation of the resistive kink instability which can prevent complete reconnection during sawtooth crashes. \cite{ORo91,Naga91} It is also important for reconnection at the dayside magnetopause \cite{Swi03,Cas07} where the interface between the incoming solar wind and the Earth's magnetosphere is characterized by a strong density gradient. Unfortunately, our assumption of a strong guide field makes our model inapplicable to the magnetopause.

The paper is organized as follows. In Sec.~\ref{sec:mod} we review the model equations. Sec.~\ref{sec:lin} is devoted to the simulation results in the linear phase and to the comparison with analytical predictions. In Sec.~\ref{sec:non} we consider the evolution of the system in the nonlinear phase. We conclude in Sec.~\ref{sec:conc}.

\section{The model}  \label{sec:mod}

We consider the two-dimensional version of the Hamiltonian gyrofluid model described in Ref.\cite{Wae09}. The model equations are 
\begin{eqnarray}
\frac{\pp n_i}{\pp t}+[ \Phi,n_i ]=0, \label{eqc1}\\
\frac{\pp n_e}{\pp t}+[\phi,n_e ]-[\psi,\lap \psi]=0,\label{eqc2}\\
\frac{\pp}{\pp t}(\psi - d_e^2 \lap \psi)+[\phi,\psi- d_e^2 \lap \psi]+\rho_s^2[\psi,n_e]=0, \label{eqc3}\\
n_e=\Gamma_0^{1/2}n_i+(\Gamma_0 -1)\phi/\rho_i^2. \label{eqc4}
\end{eqnarray}
These represent the continuity equation for the ion guiding centers, the electron continuity equation, the Ohm's law and the Poisson's equation, respectively.
In Eqs.(\ref{eqc1})-(\ref{eqc4}) $n_i$ is the ion guiding center density, $n_e$ the electron density, $\phi$ the electrostatic potential, $\psi$ the poloidal magnetic flux function, $d_e$ the electron skin depth, $\rho_s$ the sonic Larmor radius and $\Phi=\Gamma_0^{1/2}\phi$ is the gyro-averaged electrostatic potential. For our analysis the Pad\'e approximant version $\Gamma_0^{1/2}=(1-\rho_i^2 \lap/2)^{-1}$ of the gyro-average operator will be adopted, where $\rho_i$ is the ion Larmor radius. Given a Cartesian coordinate system $(x,y,z)$, we assume all the fields be translationally invariant along $z$ and we define the canonical bracket between two generic fields $f$ and $g$ by $[f,g]=\hat{z}\cdot\nabla f \times \nabla g$. We also recall that we normalize the time with respect to the Alfv\`en time and distances with respect to a magnetic equilibrium scale length $L$. Dependent variables are normalized in the following way
\benn \label{modsp}
\qquad n_i=\frac{L}{\hat{d}_i}\frac{\hat{n}_i}{n_0}, \qquad n_e=\frac{L}{\hat{d}_i}\frac{\hat{n}_e}{n_0}, \qquad \psi=\frac{\hat{A_z}}{BL}, \qquad \phi=\frac{\hat{\rho}_s^2}{L^2}\frac{L}{\hat{d}_i}\frac{e\hat{\phi}}{T_e}, \nonumber
\eenn
where carets denote dimensional variables, $d_i$ is the ion skin depth, $n_0$ a background density amplitude, $A_z$ the magnetic potential, $e$ the unit charge, $B$ a characteristic magnetic field amplitude and $T_e$ the electron temperature, which is assumed to be constant. 

\section{Numerical simulations in the linear regime}  \label{sec:lin}

We devote this section to the analysis of the numerical solutions of the model (\ref{eqc1})-(\ref{eqc4}) in the phase in which nonlinear terms are not dominant.\\ 
We solve the model equations over the domain $\{(x,y):-\pi\leq x < \pi, -a\pi\leq y < a\pi\}$, where we prescribe the value of the number $a$ at each simulation. The grid is composed by 1024$\times$128 points and double periodic boundary conditions are imposed on the field perturbations, which are the quantities that the code advances in time. Unlike Ref.~\cite{Gra10} we consider now {\it inhomogeneous} density equilibria. The simulations are then started by perturbing the equilibrium
\beq \label{equil}
n_{i_ {eq}}(x)=n_0' x, \qquad n_{e_ {eq}}(x)=n_0' x, \qquad \psi_{eq}(x)=\sum_{n=-11}^{11} a_n \exp(inx),
\eeq
where $n_0'$ is a constant and the $a_n$ are the Fourier coefficients of the function $f(x)=1/\cosh^2 x$. With respect to the simulations of Ref.~\cite{Gra10} the presence of equilibrium density gradients makes the evolution of $n_i$ no longer negligible and introduces the diamagnetic effects. A global dispersion relation for the system in the presence of diamagnetic effects was derived in Ref.~\cite{Wae09}. This showed also how the Pad\'e approximant version of the gyrofluid model was able to reproduce adequately dispersive properties of the system, such as a spectral gap in the frequencies, which are not caught by Finite Larmor Radius (FLR) models.\\
The equilibrium (\ref{equil}) is perturbed in $n_i$ with a four-cell pattern disturbance of the form $\tilde{n}_i \propto \cos(x+y/a)-\cos(x-y/a)$. The field $\phi$ is also perturbed according to (\ref{eqc4}), in such a way that the initial perturbation on $n_e$ is zero.

\begin{table}
\caption{\label{tab1}Table displaying values of linear growth rate $\gamma$ and rotation frequency $\omega$, for the dominant mode. A comparison is made between values obtained from numerical simulations ($\omega_{num}$,$\gamma_{num}$) and values ($\omega_{theor}$,$\gamma_{theor}$) predicted by the asymptotic theory of Ref.~\cite{Por91}. The agreement between numerical and analytical results is better for $\Delta'=59.9$, which is a value closer to the asymptotic regime of validity of the theory. For all cases $d_e=0.2$.}
\begin{center}
\begin{tabular}{lllllllll}
\br
$\Delta'$ & $\rho_i$ &  $\rho_s$ & $v_*$ &  $\gamma_{num}$ & $\gamma_{theor}$ & $\omega_{num}$ & $\omega_{theor}$ & Error \\
\mr
34.2& 0.4 & 0.4 & -0.16 & 0.33 & 0.44 & 0 & 0 & 24 $\%$\\
34.2& 0.4 & 0.4 & -0.4 & 0.23 & 0.37 & -0.02 & 0 & 38 $\%$\\
34.2& 0.35 & 0.2 & -0.04 & 0.28 & 0.36 & 0.03 & 0.11 & 22 $\%$\\
59.9& 0.4 & 0.4 & -0.16 & 0.27 & 0.33 & 0 & 0 & 19 $\%$\\
59.9& 0.4 & 0.4 & -0.4 & 0.21 & 0.28 & -0.01 & 0 & 27 $\%$\\
59.9& 0.35 & 0.2 & -0.04 & 0.23 & 0.27 & 0.03 & 0.08 & 15 $\%$\\
\br
\end{tabular}
\end{center}
\end{table}
If we indicate with $\chi(x,y,t)$ any of the three dynamical variables of our system, i.e. $n_i$, $n_e$ and $\psi$, then we can decompose the fields as $\chi=\chi_{eq}(x)+\tilde{\chi}(x,y,t)$. We write then the perturbation $\tilde{\chi}$ as a Fourier series truncated at $N$ modes, so that $\tilde{\chi}=\sum_{-N}^N\hat{\tilde{\chi}}_{\mathbf{k}}\exp(i \mathbf{k}\cdot \mathbf{x}+\lambda t)$, where the hat indicates the Fourier transform, $\mathbf{k}$ is the wave vector and $\lambda=\gamma - i \omega$ is a complex number having the growth rate $\gamma$ as real part and the rotation frequency $-\omega$ as imaginary part. Simulations with a relatively small amplitude of the initial perturbation, corresponding to $10^{-6}$, have been carried out, in order to obtain a long linear phase ($\sim 25$ Alfv\`en times for most of the simulations) that can be effectively investigated. In particular, we are interested in extracting from the simulations the values of $\gamma$ and $\omega$, which can permit us to observe the effects of equilibrium density gradients and finite gyro-radius, on the linear mode stabilization and island rotation frequency. We find it also of interest to compare the obtained numerical results with those predicted by the theory developed in Ref.~\cite{Por91}. The linear analysis by Porcelli refers namely to the collisionless case in the presence of finite ion temperature. Here we limit ourselves to recall the results of the theory that are relevant for our analysis.  By adopting a matching asymptotic technique, in Ref.~\cite{Por91}, the following dispersion relation, for the main mode, is obtained
\beq  \label{disp}
\lambda^2-\omega_{*e}\omega_{*i}+i\lambda(\omega_{*e}+\omega_{*i})\approx\gamma_0^2.
\eeq 
In (\ref{disp}) the electron diamagnetic frequency $\omega_{*e}$ is defined by $\omega_{*e}=-k_y c T_e n_0'/(eB n_0 v_A)$, where $v_A$ is the Alfv\`en speed based on $B$. The ion diamagnetic frequency is $\omega_{*i}=-\tau\omega_{*e}$, with $\tau=\rho_i^2/\rho_s^2$, and $\gamma_0$, for our case, reads $\gamma_0=2 k_y (2 d_e (\rho_s^2+\rho_i^2)/\pi)^{1/3}$. Splitting the dispersion relation into its real and imaginary part yields
\benn
\omega\approx\frac{\omega_{*e}}{2}(1-\tau), \label{ome}\\
\gamma^2\approx\gamma_0^2-\left[\frac{\omega_{*e}}{2}(1+\tau)\right]^2, \label{lam}
\eenn
which give us a prediction for the growth rate and the rotation frequency. Such results, however, are valid only in the limit $\rho_i \gg d_e$ and $\Delta' \gg \/\rho_i^{1/3}d_e^{2/3}$, where $\Delta'$ is the standard stability parameter for tearing modes.\\ 
Examples of theoretical and numerical values are provided in Table~\ref{tab1}. Two main cases are considered, corresponding to two different values of $\Delta'$ (which correspond to the values $a=1.5$ and $a=2$ of the parameter scaling the length of the domain along $y$). For each of these two cases the same sets of parameters $\rho_i$, $\rho_s$ and $v_*=\omega_{*e}/k_y$ are considered. From these representative examples we can see that the numerical results follow the trends predicted by the theory, for the dominant mode (the latter is $k_y=0.66$ for $\Delta'=34.2$ and $k_y=0.5$ for $\Delta'=59.9$). In particular, for fixed $\rho_i$ and $\rho_s$, the cases with $v_*=-0.4$ exhibit smaller growth rate than those with $v_*=-0.16$. This indicates the stabilizing effect operated by diamagnetic flows and corresponding to the term proportional to $\omega_{*e}^2$ in (\ref{lam}). We report also that another simulation, again with $\rho_i=\rho_s=0.4$, but with $v_*=-0.64$, showed no growth of the magnetic island, indicating the complete linear stabilization due to diamagnetic effects. For these two sets of parameters, the numerical solutions show almost no rotation frequency. On the other hand, for the case $\rho_i=0.35$, $\rho_s=0.2$ and $v_*=-0.04$, the numerical solutions show a positive value of $\omega$, which corresponds to a propagation along the direction of the ion diamagnetic drift (which in our case corresponds to the positive $y$ direction). The propagation along the ion diamagnetic drift direction is indeed also what is predicted by the theory, according to (\ref{ome}), when ions are hotter than electrons. With regard to the quantitative agreement with the theory, Table~\ref{tab1} shows that this improves when increasing $\Delta'$. In the case with larger $\Delta'$ the distances between the numerical and predicted values get shorter and we can quantify that the relative error drops from an average of $28\%$ to $18.7\%$. The reduction of the error is of course a consequence of the fact that, by increasing $\Delta'$, we get closer to the asymptotic regime in which the theory is valid.

\section{Numerical simulations in the nonlinear regime} \label{sec:non}

In this section we focus on the evolution of the system in the nonlinear phase.

\begin{figure}[h!]
\centering
\includegraphics[width=6.5cm]{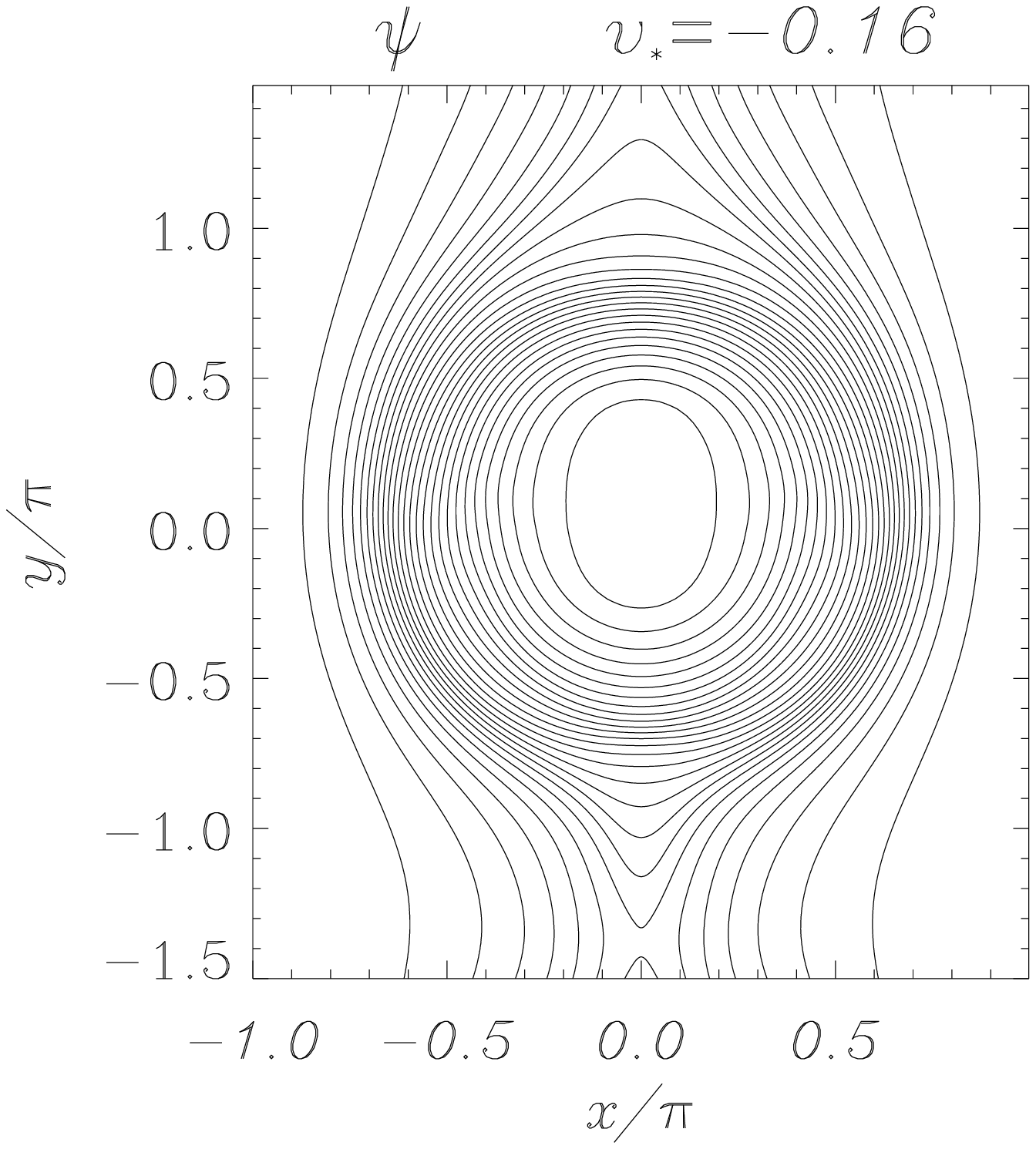}
\includegraphics[width=6.5cm]{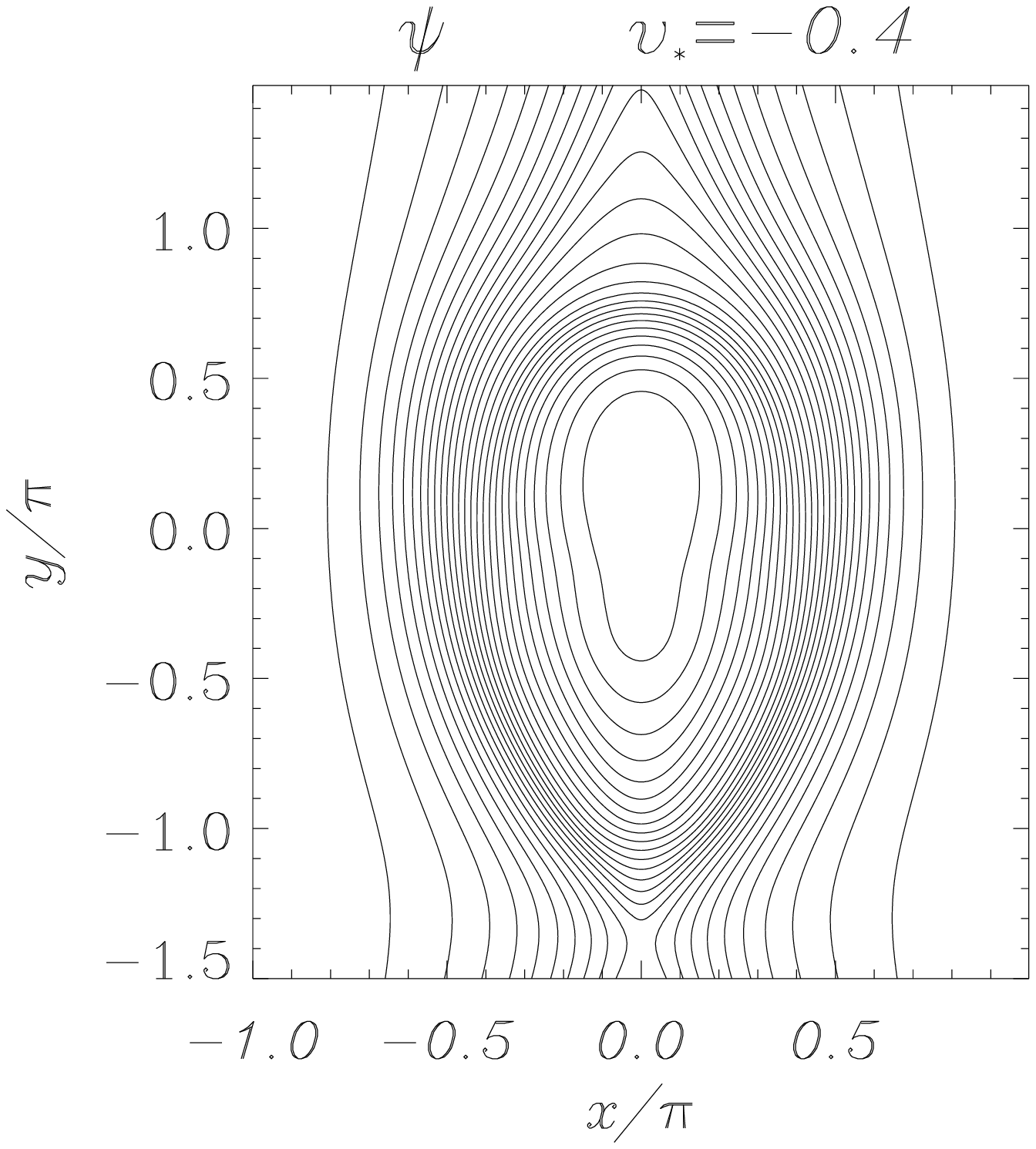}
\caption{Contour plots of $\psi$ for $v_*=-0.16$, at $t=39$ (left) and $v_*=-0.4$ at $t=50$ (right). The values of the other parameters are: $\rho_s=\rho_i=0.4$, $d_e=0.2$. The plots indicate that stronger diamagnetic effects lead to a smaller saturated island.}
\label{fig1} 
\end{figure}

Fig.~\ref{fig1} shows a comparison between contour plots of $\psi$ for $v_*=-0.16$ and $v_*=-0.4$, at times at which the two simulations reached the same phase of the dynamical evolution. From the figure we can see that the presence of stronger diamagnetic effects, apart from reducing the linear growth rate, also leads to a nonlinear island of smaller width. We recall that a nonlinear stabilizing effect of $n_0'$ was observed in a cold-ion, three-field model in cylindrical geometry \cite{Bis97}. In that context diamagnetic flows were indeed suggested to be responsible for the saturation of islands at a small width. We also note that, in spite of the fact that, in the linear phase, the islands exhibited very little rotation for these parameters, in the nonlinear regime, upward shifts of the  O-points are present, in particular in the case $v_*=-0.4$. Island propagation in the ion diamagnetic drift direction, caused either by viscosity \cite{FitzWael04a,FitzWatsWael}  or by nonlinear zonal flows, \cite{Uza10,Uza10b} has been reported in FLR simulations of visco-resistive reconnection.    \\
An effective way to look at the nonlinear dynamics of our system is to express the model in terms of its ``normal fields'', suggested by the Hamiltonian structure. We recall that the model equations (\ref{eqc1})-(\ref{eqc4}) can be rewritten as \cite{Wae09}
\begin{eqnarray}
\frac{\pp n_i}{\pp t}+[ \Phi,n_i ]=0,\\
\frac{\pp G_+}{\pp t}+[ \phi_+,G_+ ]=0,\\
\frac{\pp G_-}{\pp t}+[ \phi_-,G_- ]=0,
\end{eqnarray}
where $G_{\pm}=\psi-d_e^2\lap\psi\pm d_e\rho_s n_e$ and $\phi_{\pm}=\phi\pm(\rho_s/d_e)\psi$. The fields $n_i$ and $G_{\pm}$, which we denote as normal fields, represent natural variables for the system because, in terms of them, the model equations take the form of simple advection equations. As pointed out in a number of previous papers on reconnection based on Hamiltonian fluid models \cite{Gra01,Por02,Del05,Del06,Tas10,Tas10b}, the stream functions $\phi_{\pm}$ correspond to velocity fields that rotate in opposite directions. The fields $G_{\pm}$, advected by such  velocity fields, get then stirred in opposite directions. The stirring process causes $G_{\pm}$ to form finer and finer structures, leading to an energy cascade toward small scales. Because $n_e$ is proportional to the difference between $G_+$ and $G_-$, such cascade inevitably reflects on the electron density structures.

\begin{figure}[h!]
\centering
\includegraphics[width=6.3cm]{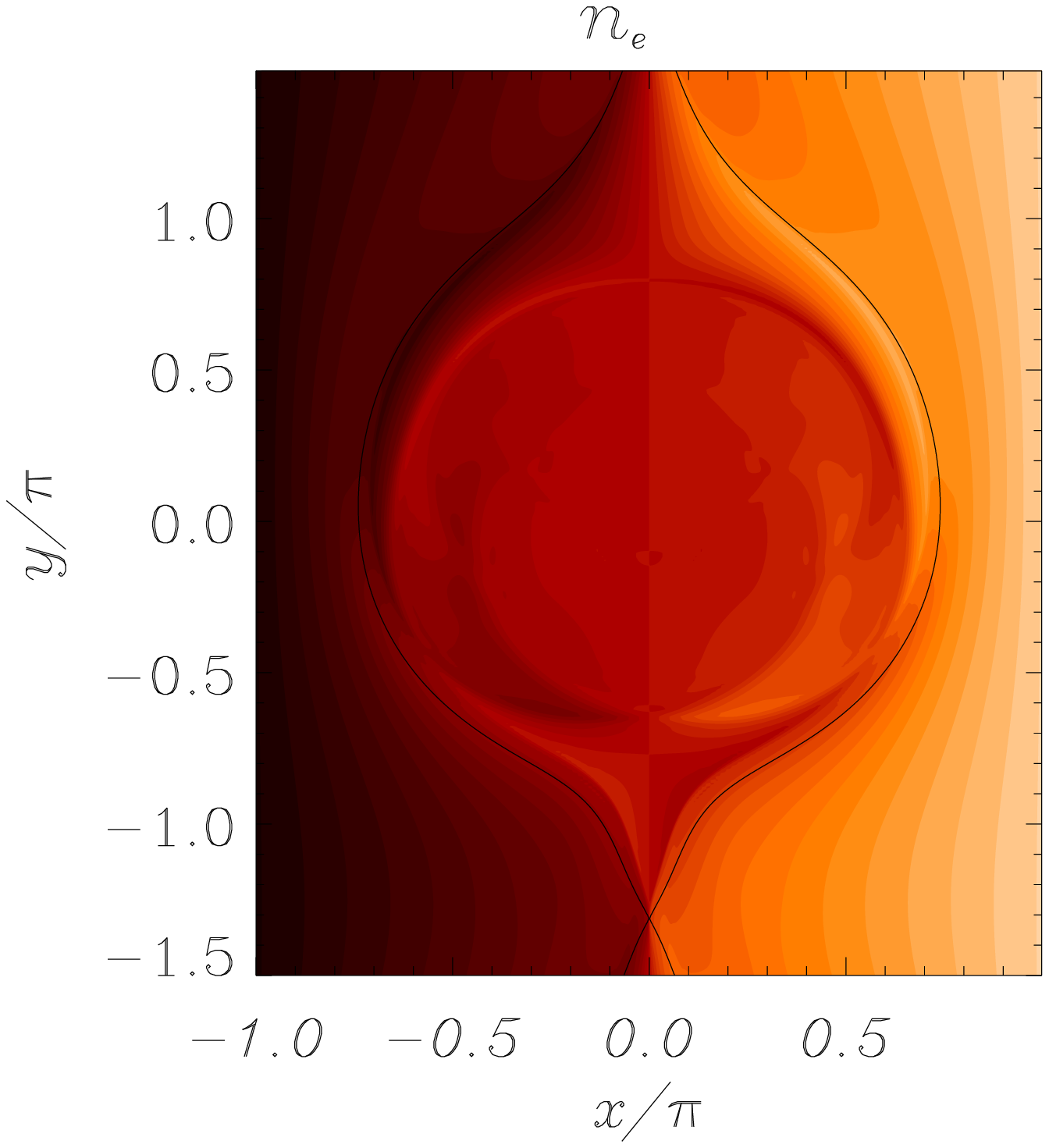}
\includegraphics[width=6.3cm]{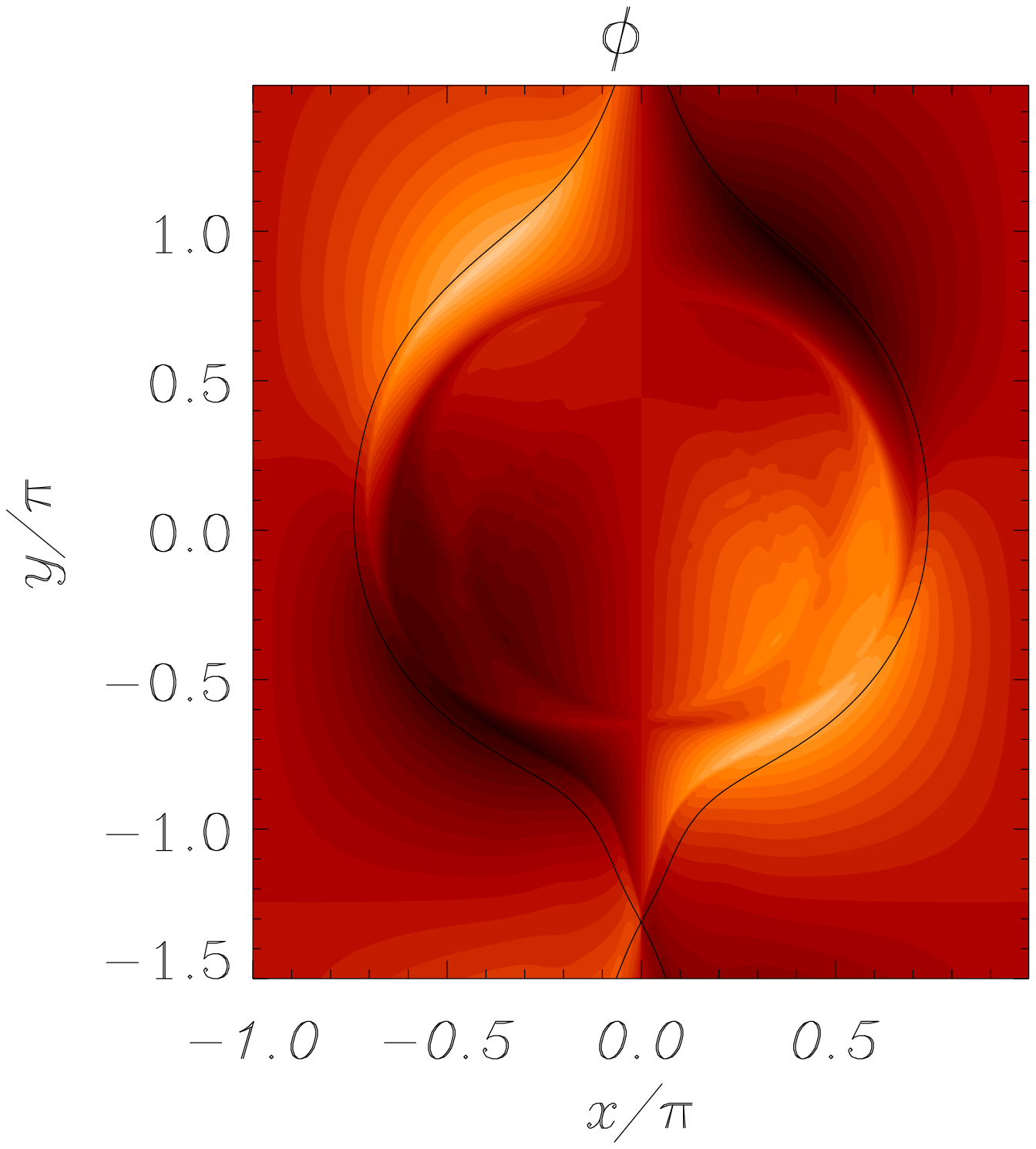}\\
\includegraphics[width=6.3cm]{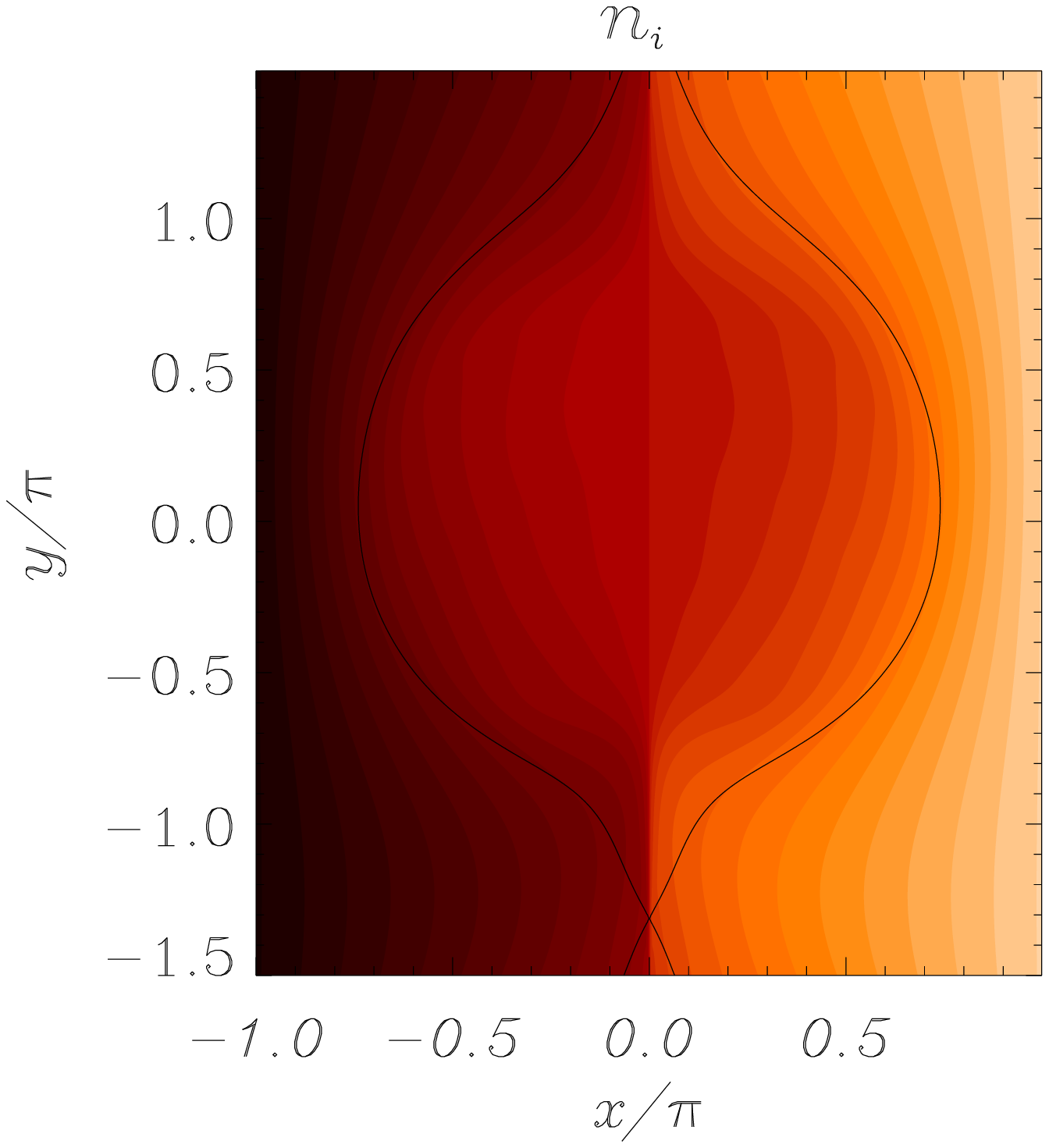}
\includegraphics[width=6.3cm]{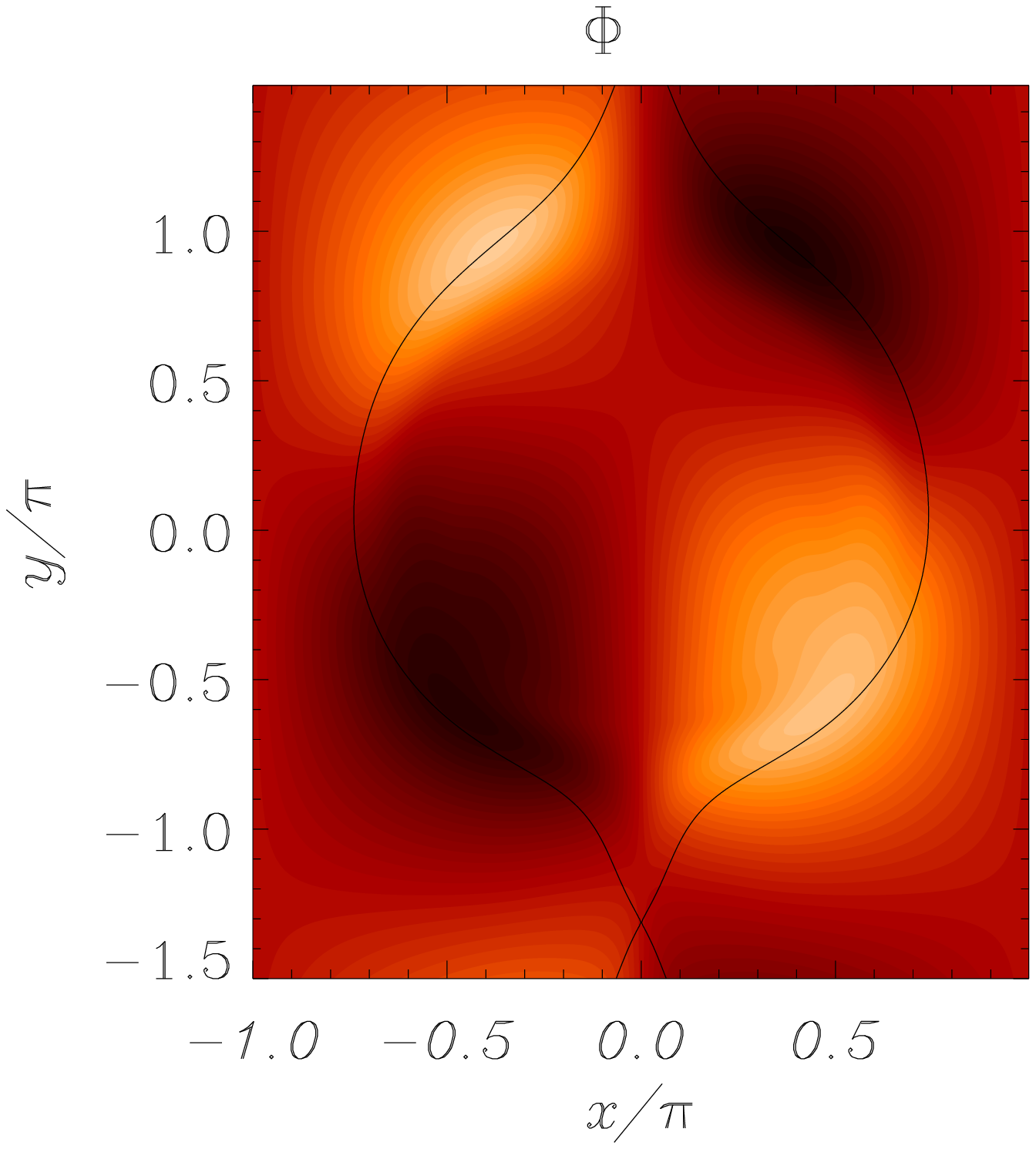}\\
\includegraphics[width=6.3cm]{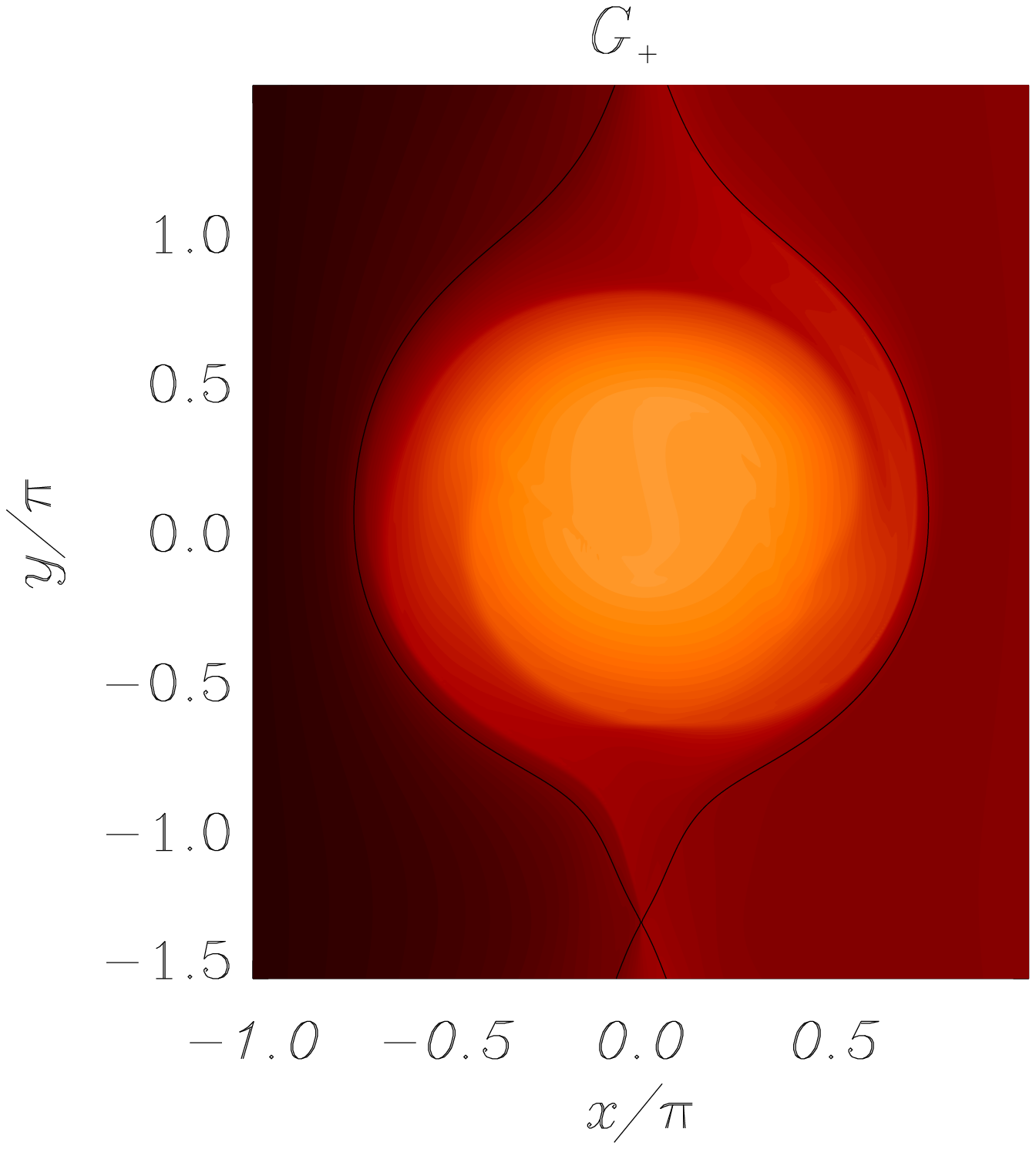}
\includegraphics[width=6.3cm]{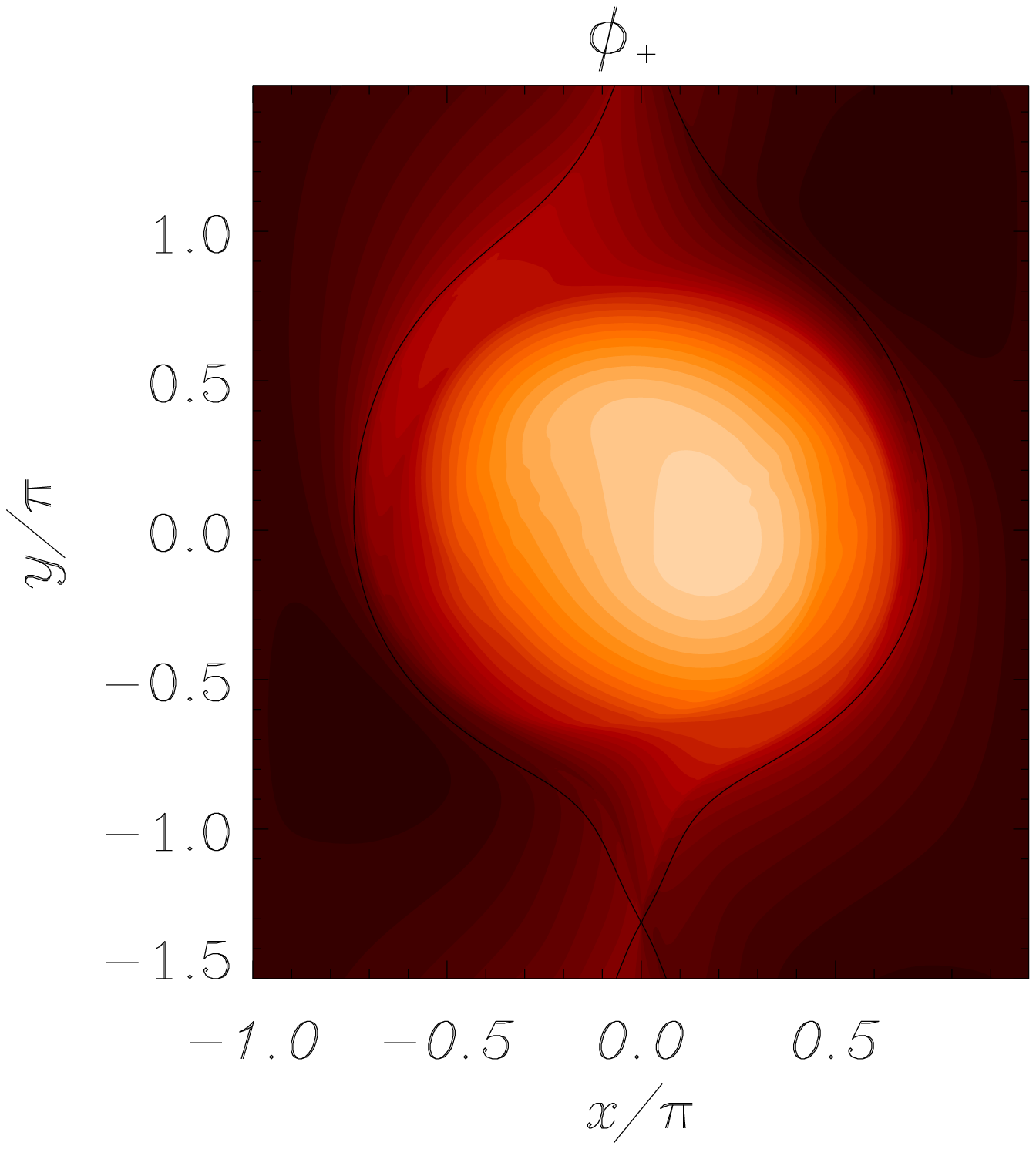}
\caption{Contour plots of $n_e$, $\phi$, $n_i$, $\Phi$, $G_+$ and $\phi_+$ at $t=19$, for $v_*=-0.16$, $\rho_i=\rho_s=0.4$ and $d_e=0.2$. The magnetic island at $t=19$ is superimposed onto each plot.}
\label{fig2}
\end{figure}

In order to investigate the nonlinear structures, we double the number of grid points in the $y$ direction, which enables us to better resolve small scale structures that form nonlinearly. We also bring up to $10^{-3}$ the amplitude of the initial perturbation.
From Fig.~\ref{fig2} indeed we see that electron density, as is typical of Hamiltonian reconnection, develops lobes around the separatrices. These emerge as the difference between the spiral arms present in $G_+$ and $G_-$ (the latter not shown here but such that $G_-(x,y,t)=G_+(-x,y,t)$). Unlike the above mentioned studies of Hamiltonian reconnection, however, the structures of Fig.~\ref{fig2} are of course not symmetric with respect to the $y=0$ axis, because of diamagnetic effects. The pattern of $\phi_+$ (and, analogously of $\phi_-(x,y,t)=-\phi_+(-x,y,t)$) are such to create thinner and shorter layers in the lower part of the magnetic island whereas wider and longer lobes are present in the upper part of the magnetic island. Inside the island, stirring of $G_{\pm}$ takes place and the initial linear density profile gets flattened.    
The ion guiding center density gradient also gets partly flattened in the center of the island, as a consequence of the advection operated by $\Phi$. Such stream function, however, has a circulation pattern quite different from that obtained by the combination of $\phi_+$ and $\phi_-$. Therefore $n_i$ does not tend to accumulate around the separatrices, unlike $n_e$. The four convective cells of the gyroaveraged electrostatic potential, on the other hand, tend to lie along the magnetic separatrices. Such behavior is also evident in the electrostatic potential $\phi$. By virtue of the relation $\phi=(1-\rho_i^2\lap/2)\Phi$, we can also infer that $\phi$ exhibits, in general, finer structures, compared to $\Phi$. We can indeed observe regions with steep gradients of electrostatic potential, around the separatrices, similarly to what happens to the electron density. Nevertheless, in this respect, it is important to recall that, for $n_0'=0$, the growth of $n_i$ was negligible, and one had $\tilde{\phi}_{\mathbf{k}}\sim -\rho_i^2 {\tilde{n_e}}_{\mathbf{k}}$, as $\rho_i\rightarrow +\infty$. This led to conclude \cite{Gra10} that the filamentation in $n_e$ implied an analogous phenomenon in $\phi$, with consequent formation of strong electric fields. In the presence of diamagnetic effects, on the other hand, Eq. (\ref{eqc4}), in the large $\rho_i$ limit, yields 
\beq
\tilde{\phi}_{\mathbf{k}}\sim-\rho_i^2 {\tilde{n_e}}_{\mathbf{k}}+\frac{2}{k^2}{\tilde{n_i}}_{\mathbf{k}}. 
\eeq
Therefore, $\phi$ is no longer proportional to $\tilde{n}_e$, but is given by the superposition of two contributions: one, related to $\tilde{n}_e$, which is responsible for the formation of steep gradients, and a second one, due to $-2\lapi \tilde{n}_i$ which, on the contrary, is relevant at large scales. 
\section{Conclusions}  \label{sec:conc}

Numerical simulations of a Hamiltonian electromagnetic gyrofluid model have been carried out and investigated. The present analysis represents a step forward, with respect to a recent related work \cite{Gra10}, because it accounts for inhomogeneous density equilibria, which lead to diamagnetic effects. In the linear phase, diamagnetic effects are shown to stabilize the growth rate of the dominant mode and induce a propagation of the magnetic island. Simulations also indicate that the direction of propagation tends to move from that of the electron to that of the ion diamagnetic drift, as the ion temperature is increased. Such features are in agreement with what is predicted by the linear theory of Ref.\cite{Por91} and the quantitative agreement improves as the values of parameters approach the asymptotic regime of validity of the theory. In the nonlinear phase, stronger equilibrium gradients tend to yield smaller ``saturated'' islands, as already observed in Ref.\cite{Bis97}. We note also a different distribution of the ion guiding center density with respect to the electron density, with the former retaining a four-cell pattern, whereas the latter concentrates around separatrices, although in an asymmetric way, also with respect to the $y=0$ axis, as a consequence of the diamagnetic terms. Compared to the homogeneous equilibrium case \cite{Gra10}, the potential possesses now a smoother contribution, due to the finite $n_i$ fluctuations. We finally indicate the derivation of a refined version of the linear theory, and the extension of the model to include parallel dynamics, as subjects of work in progress. 

\ack
This work was supported by the European Community under the contracts of Association between EURATOM and ENEA and between EURATOM, CEA, and the French Research Federation for fusion studies. FW was supported in part by a fellowship from the Katholieke Universiteit Leuven and by the US Department of Energy under contract DE-FG02-04ER-54742. The views and opinions expressed herein do not necessarily reflect those of the European Commission. Financial support was also received from the Agence Nationale de la Recherche (ANR EGYPT).

\section*{References}

\bibliographystyle{iopart-num}
\bibliography{Tassi}

\end{document}